\documentclass[10pt]{article}

\usepackage{amsmath}
\usepackage{amssymb}

\usepackage{graphicx}

\usepackage{cite}

\usepackage{color}

\usepackage{graphicx}
\usepackage{cite} 
\usepackage{url}  
\usepackage{ifthen}  
\usepackage{multicol}   
\usepackage{multirow}
\usepackage{subfigure}
\usepackage{epsfig}
\usepackage{epstopdf}
\usepackage[labelfont=bf,labelsep=period,justification=raggedright]{caption}

\makeatletter
\renewcommand{\@biblabel}[1]{\quad#1.}
\makeatother

\date{}

\begin{document}

\begin{flushleft}
{\Large
\textbf{Comparing Stochastic Differential Equations and Agent-Based Modelling and Simulation for Early-stage Cancer}
}
\\
Grazziela P Figueredo$^{1,\ast}$,
Peer-Olaf Siebers$^{1}$,
Markus R Owen$^{2}$,
Jenna Reps$^{1}$,
Uwe Aickelin$^{1}$
\\
\bf{1} Intelligent Modelling and Analysis Research Group, School of Computer Science, The University of Nottingham, NG8 1BB, UK
\\
\bf{2} Centre for Mathematical Medicine and Biology, School of Mathematical Sciences, The University of Nottingham, NG7 2RD, UK
\\
$\ast$ E-mail: grazziela.figueredo@nottingham.ac.uk
\end{flushleft}

\section*{Abstract}

There is great potential to be explored regarding the use of agent-based modelling and simulation as an alternative paradigm to investigate early-stage cancer interactions with the immune system. It does not suffer from some limitations of ordinary differential equation models, such as the lack of stochasticity, representation of individual behaviours rather than aggregates and individual memory. In this paper we investigate the potential contribution of agent-based modelling and simulation when contrasted with stochastic versions of ODE models using early-stage cancer examples. We seek answers to the following questions: (1) Does this new stochastic formulation produce similar results to the agent-based version? (2) Can these methods be used interchangeably? (3) Do agent-based models outcomes reveal any benefit when compared to the Gillespie results? To answer these research questions we investigate three well-established mathematical models describing interactions between tumour cells and immune elements. These case studies were re-conceptualised under an agent-based perspective and also converted to the Gillespie algorithm formulation. Our interest in this work, therefore, is to establish a methodological discussion regarding the usability of different simulation approaches, rather than provide further biological insights into the investigated case studies. Our results show that it is possible to obtain equivalent models that implement the same mechanisms; however, the incapacity of the Gillespie algorithm to retain individual memory of past events affects the similarity of some results. Furthermore, the emergent behaviour of ABMS produces extra patters of behaviour in the system, which was not obtained by the Gillespie algorithm.

\section*{Introduction}

In previous work, three case studies using established mathematical models of immune interactions with early-stage cancer were considered in order to investigate the additional contribution of ABMS to ODE models simulation~\cite{Figueredo2013a}. These case studies were re-conceptualised under an agent-based perspective and the simulation results were compared with those from the ODE models. Our results showed that, apart from the well known differences between these approaches (as those outlined for example, in Schieritz and Milling~\cite{Schieritz:2003}), further insight from using ABMS was obtained, such as extra population patterns of behaviour.

In this work we apply the Gillespie algorithm~\cite{Gillespie:1976,Gillespie:1977}, which is a variation of the Monte Carlo method, to create stochastic versions of the original ODE models investigated in~\cite{Figueredo2013a}. We aim to reproduce the variability embedded in the ABMS systems to the mathematical formulation and verify whether results resemble each other. In addition, due to the fact that the Gillespie algorithm also regards integer quantities for their elements, we hope that this method overcomes some differences observed when comparing atomic agents represented in the ABMS with possible fractions of elements observable in the ODE results.

To the best of our knowledge, current literature regarding the direct comparison of the Gillespie algorithm and ABMS is scarce. We want therefore to answer research questions such as: (1) Does this new stochastic formulation produce similar results to ABMS? (2) Can these methods be used interchangeably for our case studies? (3) Does the stochastic model implemented using the Gillespie algorithm also find the extra patterns revealed by the ABMS? We aim to establish a methodological discussion regarding the benefits of each approach for biological simulation, rather than provide further insights into the biological aspects of the problems studied. We therefore intend to compare the dynamics of each approach and observe the outcomes produced over time. We hope that this study provides further insights into the potential usability and contribution of ABMS to systems simulation.

\subsection*{Case Studies}

The case studies used for our comparison regard models with different population sizes, varying modelling effort and model complexity. We hope that by tackling problems with different characteristics, a more robust analysis of our experiments is performed. The features of each case study are shown in Table~\ref{Tab:CaseStudies}.

The first case study considered is based on an ODE model involving interactions between generic tumour and effector cells. The second case study adds to the previous model the influence of the IL-2 cytokine molecules in the immune responses. The third case study comprises a model of interactions between effector cells, tumour cells, and IL-2 and TGF-$\beta$ molecules. For all case studies, three approaches are presented: the original mathematical model, its conversion into the Gillespie algorithm model and the ABMS model. To answer our research questions, the Gillespie and ABMS approaches outcomes are compared. Our results show that for most cases the Gillespie algorithm does not produce outcomes statistically similar to the ABMS. In addition, Gillespie is incapable to reproduce the extreme patterns observed in the ABMS outcomes for the last case study.

The remainder of this paper is organized as follows. The next section introduces the literature review comparing stochastic ODE models and ABMS for different simulation domains. First, we show general work that has been carried out in areas such as economics and operations research, and then we focus on research concerned with the comparison for immunological problems. Finally, we discuss gaps in the literature regarding cancer research. In the following section we introduce our Gillespie and agent-based modelling development processes and the methods used for conducting the experimentation. Subsequently we present our case studies, comparison results and discussions. In the final section we draw our overall conclusions and outline future research opportunities.

\section*{Related Work}
\label{RelatedWork}

Current {\it in-silico} approaches used in early-stage cancer research include computational simulation of compartmental models, individual-based models and rule-based models. Compartmental models adopt an aggregate representation of the elements in the system. They include deterministic methods such as ordinary differential equation (ODE) models, system dynamics (SD) models and partial differential equation (PDE) models. These models have been largely employed in the study of dynamics between cancer cells and tumour cells~\cite{Arciero:2004,Eftimie2010}, therapies for cancer~\cite{Owen:2011:Cancer-Res:21363914}, tumour responses to low levels of nutrients~\cite{Kuznetsov:1994,Kirschner:1998,alarcon2004MME,alarcon2006MMT} and tumour vascularization\cite{alarcon2003CAM,Owen2009AVRNCT}. Although these models have been very useful to understand and uncover various phenomena, they present several limitations. For instance, they do not encompass emergent behaviour and stochasticity. In addition, it is difficult to keep a record of individual behaviour and memory over the simulation course~\cite{Borshchev:2004,Louzoun:2007}. Stochastic compartmental models include Monte Carlo simulation models, which are computational algorithms that perform random sampling to obtain numerical results~\cite{Metropolis:1949}. Amongst others, they are useful for simulating biological systems, such as cellular interactions and the dynamics of infectious diseases~\cite{Ball:2009}. As these methods rely on stochastic process to produce their outputs, they overcome some of the limitations of the deterministic compartmental models, as they allow for variability of outcomes. The individuals in these models, however, do not have any sort of memory of past events. Rule-based models are a relatively new research area mostly focused on modelling and simulating biochemical reactions, molecular interactions and cellular signalling. The literature regarding the application of rule-based models to interactions between the immune system and cancer cells, however, is scarce. Individual-based models, or agent-based modelling and simulation (ABMS), relax the aggregation assumptions present in compartmental approaches and allow for the observation of the behavior of the single cells or molecules involved in the system. This approach has also been applied to early-stage cancer research~\cite{Figueredo2013a,Figueredo2013b}.

The differences between deterministic compartmental models and individual-based models are well known in operations research~\cite{Scholl:2001a,Pourdehnad:2002,Schieritz:2002,Schieritz:2003} and have also been studied in epidemiology~\cite{Sterman:2008,Jaffry:2008} and system's biology~\cite{Wayne:2004,Figueredo:2010,Figueredo:2011}. Deterministic compartmental models assume continuous values for the individuals in the system, whereas in ABMS individual agents are represented. This peculiarity of each approach highly impacts the simulation results similarity depending on the size of the populations~\cite{Figueredo2013a}. There is still however the need for further investigations between the interchangeable use of some Monte Carlo methods and ABMS.

\subsection*{Approaches Comparison}

As mentioned previously, there are few studies that compare the Gillespie algorithm with ABMS. Most of these studies regard research in economic models and immunology. To the best of our knowledge there is no literature regarding the direct comparison of these methods to early-stage interactions between the immune system and tumour cells. This section describes relevant researches in several areas, which provided further insights into the gaps in the current literature and the research questions addressed in this paper.

There are a few attempts of re-conceptualizing agent-based models into simpler stochastic models of complex systems in economics. For instance, Daniunas {\it et al.}~\cite{Kononovicius02agent-basedversus} start from simple models with established agent-based versions (which they named ``the model's microscopic version'') and try to obtain an equivalent macroscopic behavior. They consider microscopic and macroscopic versions of the herding model proposed by Kirman~\cite{Kirman:1993} and the diffusion of new products, proposed by Bass in~\cite{Bass:1969}. They conclude that such simple models are easily replicated in a stochastic environment. In addition, the authors state that for the economics field, only very general models, such as those studied in their article, have well established agent-based versions and can be described by stochastic or ordinary differential equations. However, as the complexity of the microscopic environment increases, it becomes challenging to obtain resembling results with stochastic simulations and further developments need to be pursuit.  Furthermore, the authors debate that the ambiguity present in the microscopic description in complex systems {\it is an objective obstacle for quantitative modeling} and needs further studying.

Stracquadanio {\it et al.}~\cite{Umeton:2011} investigate the contributions of ABMS and the Gillespie method for immune modelling. The authors, however, do not apply both methods to the same problem. Instead, for the first approach, they chose to investigate a large-scale model involving interactions of immune cells and molecules. This model's objective was to simulate the immune elements interplay over time. For the Gillespie approach, the authors investigate a stochastic viral infection model. The authors point out three factors that play a major role in the modeling outcome when comparing ABMS and Gillespie: simulation time, model precision and accuracy, and model applicability. Regarding time, the authors state that stochastic models implemented with the Gillespie algorithm are preferred. On the other hand, ABMS permits more control over simulation runtime as it keeps record of the behavior of each single entity involved in the system. Regarding applicability, the authors argue that traditional Gillespie methods do not account for spatial information, which can be detrimental to the model accuracy given the fact that many immune interactions occur within specific spatial regions of the simulation environment.

Karkutla~\cite{Raja:2010} compares two biological simulators: GridCell, which is a stochastic tool based on Gillespie's, and his new developed ABMSim, which is a simulation tool based on ABMS. GridCell was developed to overcome the issues in traditional Gillespie's, as pointed out by Stracquadanio {\it et al.}~\cite{Umeton:2011}. It is a stochastic tool able to tackle non-homogeneity effectively by addressing issues of crowding and localization. In GridCell, however, the problem of tracking individual behaviour and determining particular characteristics to each element still exists. ABMSim was therefore developed to overcome these issues. GridCell has been compared with ABMSim qualitatively and quantitatively and the two tools have produced similar results in the author experiments.

In our work we further study the differences between the approaches outcomes and investigate whether under different problem characteristics for early-stage cancer we still obtain similar results. In the next section we introduce the methodology used to conduct our investigations.

\section*{Materials and Methods}
\label{Methodology}

This section introduces the research methodology used for the development of our simulation models and for the experimentation performed in the following sections. As mentioned previously, our investigations concern the use of three cases to answer research questions regarding the application of the Gillespie algorithm and ABMS interchangeably for early-stage cancer models. For each case study, there is a well established ODE model from the literature and its correspondent agent-based model, that we previously developed in~\cite{Figueredo2013a}. The ODE model simulations are implemented using the ODE solver module from MATLAB (2011).

The Gillespie algorithm is implemented by the direct conversion of the original mathematical equations into reactions and simulating them under the COPASI $4.8$ $(Build 35)$ simulator environment. The method used for the stochastic simulations is the Gillespie algorithm adapted using the next reaction method~\cite{Bruk:2000}, with interval sizes of $0.1$, integration interval between $0$ and $1$ and maximum internal steps of $10^6$.

ABMS is a modelling and simulation technique that employs autonomous agents that interact with each other. The agents' behaviour is described by rules that determines how they learn, interact and adapt. The overall system behaviour is given by the agents individual dynamics as well as their interactions. Our agent-based models are implemented using the $AnyLogic^{TM}$ 6.5 educational version (XJ Technologies 2010)~\cite{AnyLogic:2010}. This approach is developed by using state charts and tables containing each agent description. The state charts show the different possible states of an entity and define the events that cause a transition from one state to another. In order to facilitate the understanding of the agent-based model,
we reproduce here the models developments, which were based on~\cite{Figueredo2013a}~\footnote{The ABMS and Gillespie models are available for download in http://anytips.cs.nott.ac.uk/wiki/index.php/Resources}.

\subsection*{Methodology for Results Comparison}

As the Gillespie and ABMS are both stochastic simulation methods, we ran five hundred replications for each case study and calculated the mean values for the outputs. For all approaches, the rates (for cellular death, birth, etc.) employed were the same as those established by the mathematical model.

In addition, in order to investigate any statically significant differences between the ABMS and Gillespie techniques for the case studies, we implement a mixed effect model. This is a type of regression that considers both fixed and random effects.  This method accounts for correlation caused by repeating the measure over time (i.e., the tumour cell count is correlated over time for each simulation run). The mixed effect analysis was implemented in the programming language R using the package NLME \cite{nlme}.

As a mixed effect model requires finding parameters for a regression model, it is not suitable when considering the whole time period. This is because in cases 2 and 3 the tumour dynamics has a damping oscillation and the function describing this dynamics is unknown (see pages~\pageref{sec:ResultsCase2} and~\pageref{sec:ResultsCase3}).  Instead, the sequence of local maxima and minima are used. It can been seen that these are converging and any statistical deviation between these sequences for the different simulation techniques indicate differences between the output of the techniques.  If the simulations from the ABMS and the Gillespie technique come from the same distribution, then there would be no statistical difference between the maxima and minima over time. Therefore, we investigate two null hypotheses. The first is that the function of local maxima is the same for the ABMS and the Gillespie algorithm simulations. And the second is that the function of local minima is the same for the ABMS and the Gillespie algorithm simulations. We use a 1\% significance level.

There is no standard technique for estimating the required sample size for non-linear mixed effect models for a defined power when the measure of effect is known \cite{kang2004} .  Therefore, the simulations are run 500 times as this will increase the statistical power and increase the probability of a true positive in the statistical analysis.  A false negative is still possible if there is only a small effect size, but if the effect is small, it is of less interest.

\subsection*{Case 1: Interactions between Tumour Cells and Generic Effector Cells}
\label{Case1}

The first case considers tumour cells growth and their interactions with general immune effector cells, as defined in~\cite{Kuznetsov:1994}. According to the model, effector cells search and kill the tumour cells inside the organism. They proliferate proportionally to the number of existing tumour cells. As the quantities of effector cells increase, their capacity of eliminating tumour cells is augmented. Immune cells proliferate and die per apoptosis, which is a programmed cellular death. In the model, cancer treatment is also considered and it consists of injections of new effector cells in the organism.

Mathematically, the interactions between tumour cells and immune effector cells are defined as follows~\cite{Kuznetsov:1994}:

\begin{equation}
\frac{dT}{dt} = Tf(T) - d_T(T,E)
\end{equation}

\begin{equation}
\frac{dE}{dt} = p_E(T,E) - d_E(T,E) - a_E(E) + \Phi(T)
\end{equation}

where

\begin{itemize}
\item $T$ is the number of tumour cells,
\item $E$ is the number of effector cells,
\item $f(T)$ is the growth of tumour cells,
\item $d_T(T,E)$ is the number of tumour cells killed by effector cells,
\item $p_E(T,E)$ is the proliferation of effector cells,
\item $d_E(T,E)$ is the death of effector cells when fighting tumour cells,
\item $a_E(E)$ is the death (apoptosis) of effector cells,
\item $\Phi(T)$ is the treatment or influx of cells.
\end{itemize}

The Kuznetsov model~\cite{Kuznetsov:1994} defines the functions $f(T)$, $d_T(T,E)$, $p_E(E,T)$, $d_E(E,T)$, $a_E(E)$ and $\Phi(t)$ as shown below:

\begin{equation}
f(T) = a(1-bT)
\end{equation}

\begin{equation}
d_T(T,E)=nTE
\end{equation}

\begin{equation}
p_E(E,T)= \frac{pTE}{g+T}
\end{equation}

\begin{equation}
d_E(E,T)=mTE
\end{equation}

\begin{equation}
a_E(E)= dE
\end{equation}

\begin{equation}
\Phi(t) = s
\end{equation}

Table~\ref{Tab:GillespieCase1} shows the mathematical equations converted into reactions and their respective rate laws per cell.

\begin{figure}[!htpb]
\centering
\subfigure[Tumour cell agent]{
   \includegraphics[scale =0.54] {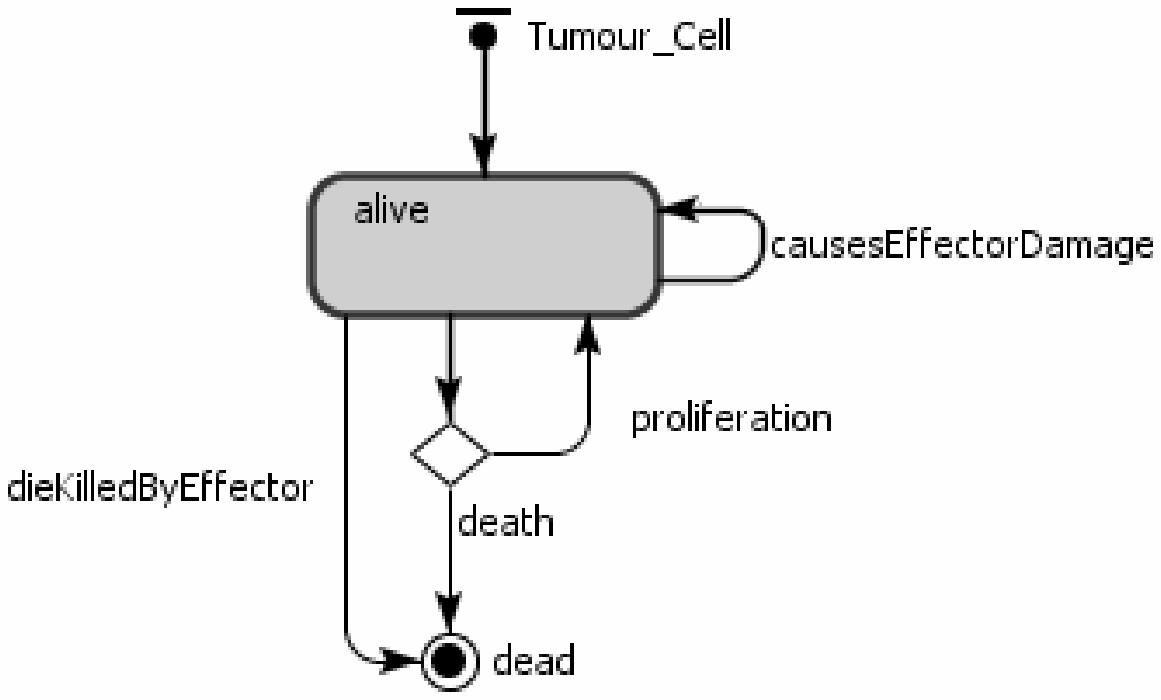}
 }
\qquad
 \subfigure[Effector cell agent]{
   \includegraphics[scale =0.54] {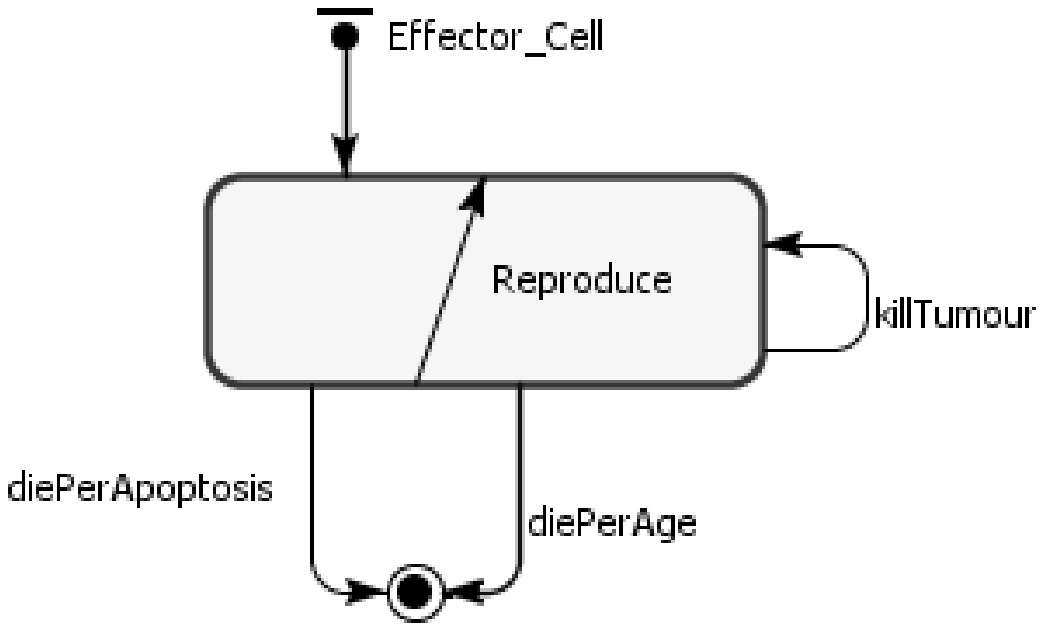}
 }
 \caption{{\bf ABMS state charts for case 1}}
 \label{fig:CaseStudy1:ABS}
\end{figure}

In the agent-based model there are two classes of agents, the tumor cells and the effector cells, as described in~\cite{Figueredo2013a}.
Table~\ref{Tab:Case1:AgentsBehaviours} shows the parameters and behaviours corresponding to each agent state. For our agents, state charts are used to represent the different states each entity is in. In addition, transitions are used to indicate how the agents move from one state to another. Events are also employed and they indicate that certain actions are scheduled to occur in the course of the simulation, such as injection of treatment. The state chart representing the tumour cells is shown in Figure~\ref{fig:CaseStudy1:ABS}(a), in which an agent proliferates, dies with age or is killed by effector cells. In addition,  tumour cells contribute to damage to effector cells, according to the same rate as defined by the mathematical model (Table~\ref{Tab:TumourEffectorTransitionRatesCalculations1}). Figure~\ref{fig:CaseStudy1:ABS}(b) shows the state chart for the effector cells. In the figure, the cell is either alive or dead by age or apoptosis. While the cell is alive, it is also able to kill tumour cells and proliferate. In the transition rate calculations, the variable $TotalTumourCells$ corresponds to the total number of tumour cell agents; and the variable $TotalEffectorCells$ is the total number of effector cell agents. In the simulation model, apart from the agents, there is also an event -- namely, treatment -- which produces new effector cells with a rate defined by the parameter $s$.

\subsubsection*{Experimental Design for the Simulations}

Similarly to the experiments from~\cite{Figueredo2013a}, four scenarios are investigated. The scenarios have different rates for the death of tumour cells (defined by parameter $b$), effector cells apoptosis (defined by parameter $d$) and different treatments (parameter $s$). The values for these parameters are obtained from \cite{Eftimie2010} (Table~\ref{Tab:CaseStudy1:TumourEffectorsParameters}). In the first three scenarios, cancer treatment is considered, while the fourth case does not consider any treatment. The simulations for the ABMS and the Gillespie algorithm are run five hundred times and the mean values are displayed as results.

\subsubsection*{Results and Discussion}
\label{sec:ResultsCase1}

\begin{figure}[!ht]
  \centering
  \resizebox{16.5cm}{!}{\includegraphics{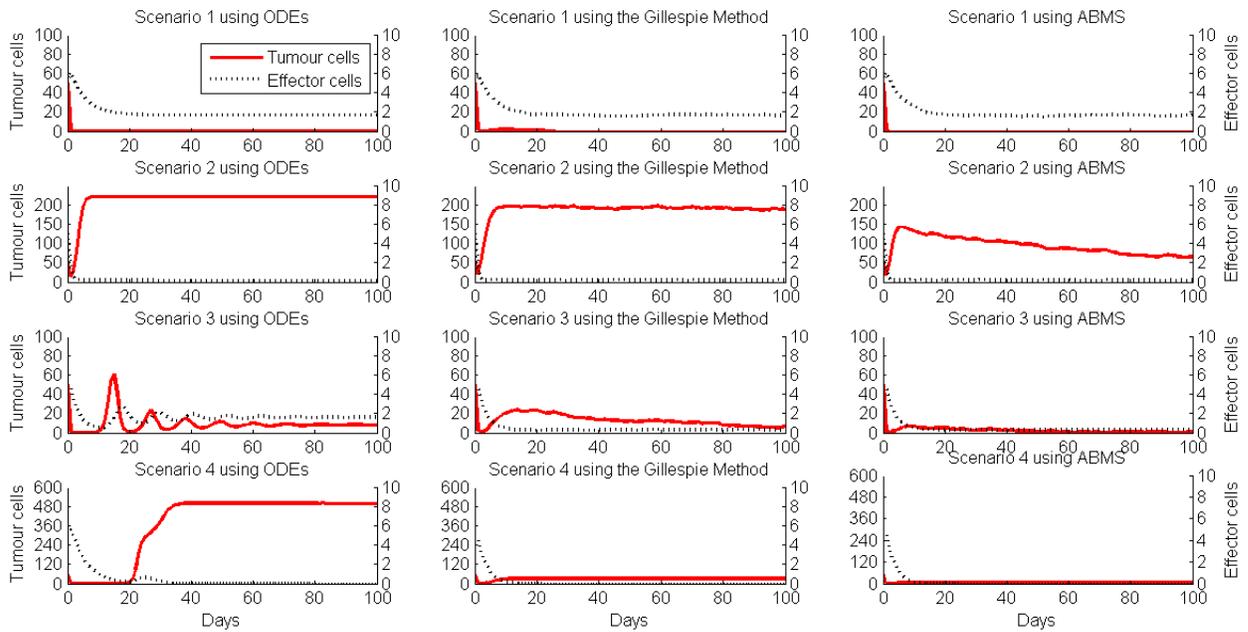}}
  \caption{{\bf Results for the first case study}}
  \label{fig:ResultsCase1}
\end{figure}

Figure~\ref{fig:ResultsCase1} shows the results of our experiments. In the first column we display the results from the ODE model for guidance. The second column shows the results from the Gillespie algorithm and the third column presents the ABMS results. Each row of the figure represents a different scenario.

Results for Scenario 1 appear similar for the three approaches, although the effector cells curve from the ABMS show more variability. To evaluate whether the results are significantly different for the two simulation methods, we apply a mixed effect model.  The null hypothesis is that there is no significant difference between the methods (and therefore there will be no significant fixed effect for the method type).  We use a 1\% significance level. We are testing the similarity for the population of effector cells.

The effector cells follow a dynamic similar to 1/x for the time between 1 and 100:
\begin{equation}
f(t) = \frac{5}{a \times (t+b)}
\end{equation}

We apply a mixed-effect model where the simulation run is considered to have a random effect on the parameter $a$ and the simulation method has a fixed effect on $a$ and $b$. The results are presented in Table \ref{tab:statsCase1}. At a 1\% significance level the results of the two techniques are significantly different as the p-values for the fixed effect of the method on the parameters are less than $0.1$.

We believe that the variability observed in the ABMS  and Gillespie curves, given their stochasticity, also influenced the statistical test results. The number of effector cells for all simulations follow a similar pattern, although the similarity hypothesis was rejected.  This variability of Gillespie and ABMS is very evident with regards to the effector cells population as the size of the populations involved in the first scenario is relatively small, which increases the impacts of stochasticity in the outcomes.

Results for scenario 2 are shown in the second row of Figure~\ref{fig:ResultsCase1}. The outcomes seem fairly different. By observing the ODE results, during about the first ten days, the tumour cells decrease and then grow up to a value of about 240 cells, subsequently reaching a steady-state. This initial decrease is also observed in both Gillespie and ABMS curves. However, only the Gillespie method shows a similar increase in the numbers of tumour cells when compared to the ODEs. Similarly to the previous scenario, the Gillespie and ABMS simulation curves present an erratic behaviour throughout the simulation days. There is, however, an unexpected decay of tumour cells over time in the ABMS simulation, which does not happen in the Gillespie outcome. We believe the difference observed in the ABMS is due to the the individual characteristics of the agents and their growth/death rates attributed to their instantiation. While both ODEs and Gillespie are compartmental models and therefore they apply the model rates to the cells population, ABMS on the contrary, employs these rates in an individual basis. As the death rates of the tumour cells agents are defined according to the mathematical model, when the tumour cell population grows, the newborn tumour cells have higher death probabilities, which leads to a considerable number of cells dying out. This indicates that the individual behaviour of cells can lead to a more chaotic behaviour when compared to the aggregate view observed in the compartmental simulation.

For scenarios 3 and 4, shown in third and fourth rows of Figure~\ref{fig:ResultsCase1}, respectively, the results for the three approaches differ completely. The differences are even more evident for the tumour cells outcomes. The ODEs results for scenario 3 reveal that tumour cells decreased as effector cells increased, following a predator-prey trend curve. For the ABMS, however, the number of effector cells decreased until a value close to zero was reached, while the tumour cells numbers were very different from those in the ODEs results. The ODE pattern noticed was possible given its continuous character. In the ODE simulation outcome curve for the effector cells it is therefore possible to observe, for instance,  that after sixty days the number of effector cells ranges between one and two. These values could not be reflected in the ABMS simulation, as it deals with integer values. Similarly, the Gillespie approach outcomes did not resemble those from the ODE model. There is more variability in the tumour cells curve than in the ABMS outcomes, although the number of tumour cells also reaches zero after around sixty days.

In the fourth scenario, although effector cells appear to decay in a similar trend for both approaches, the results for tumour cells vary largely. In the ODE simulation, the numbers of effector cells reached a value close to zero after twenty days and then increased to a value smaller than one. For the ABMS simulation, however, these cells reached zero and never increased again. For the Gillespie model results, a similar pattern as that from the ABMS model occurs, although there seems to be less variance in the outcome curve. In addition, the mean numbers for tumour cells for the Gillespie approach seem smaller that those observed in the ABMS.

\subsubsection*{Summary}

An outcome comparison between an ABMS and a Gillespie algorithm model was performed for case study 1. We considered an ODE model of tumour cells growth and their interactions with general immune effector cells as the baseline for results validation. Four scenarios considering small population numbers were investigated and results from ABMS and Gillespie were different for both populations for all scenarios. Furthermore, both approaches differed largely from the original mathematical outcomes. These results indicate that, for this case study, the stochasticity applied to the population as a whole when compared to that applied to the individual has a higher impact given the small population sizes. The result analysis also reveals that conceptualizing the stochastic approaches from the mathematical equations does not always produce statistically similar outputs.

\subsection*{Case 2: Interactions Between Tumour Cells, Effector Cells and Cytokines IL-2}
\label{Case2}

Case two regards the interactions between tumour cells, effector cells and the cytokine IL-2. It extends the previous study as it considers IL-2 as molecules mediating the immune response towards tumour cells. These molecules interfere in the proliferation of effector cells, which occurs proportionally to the number of tumour cells in the system. For this case, there are two types of treatment, the injection of effector cells or the addition of cytokines.

The mathematical model used in case 2 is obtained from~\cite{Kirschner:1998}. The model's equations described bellow illustrate the non-spatial dynamics between effector cells (E), tumour cells (T) and the cytokine IL-2 ($I_L$):

\begin{equation}
\label{Eq:cap:CaseStudy2:Case2:1}
 \frac{dE}{dt} = cT - \mu_2E + \frac{p_1EI_L}{g_1+I_L} + s1
\end{equation}

Equation~\ref{Eq:cap:CaseStudy2:Case2:1} describes the rate of change for the effector cell population E~\cite{Kirschner:1998}. Effector cells grow based on recruitment ($cT$) and proliferation ($\frac{p_1EI_L}{g_1+I_L}$). The parameter $c$ represents the antigenicity of the tumour cells (T)~\cite{Kirschner:1998,Arciero:2004}. $\mu_2$ is the death rate of the
effector cells.  $p_1$ and $g_1$ are parameters used to calibrate the recruitment of effector cells and $s1$ is the treatment that will boost the number of effector cells.

\begin{equation}
\label{Eq:cap:CaseStudy2:Case2:2}
 \frac{dT}{dt} = a(1 - bT) - \frac{a_aET}{g_2 + T}
\end{equation}

Equation~\ref{Eq:cap:CaseStudy2:Case2:2} describes the changes that occur in the tumour cell population T over time. The term $a(1 - bT)$ represents the logistic growth of T ($a$ and $b$ are parameters that define how the tumour cells will grow) and $\frac{a_aET}{g_2 + T}$ is the number of tumour cells killed by effector cells. $a_a$ and $g_2$ are parameters to adjust the model.

\begin{equation}
\label{Eq:cap:CaseStudy2:Case2:3}
 \frac{dI_L}{dt} = \frac{p_2ET}{g_3 + T} - \mu_3I_L + s2
\end{equation}

The IL-2 population dynamics is described by Equation~\ref{Eq:cap:CaseStudy2:Case2:3}. $\frac{p_2ET}{g_3 + T}$ determines IL-2 production using parameters $p_2$ and $g_3$. $\mu_3$ is the IL-2 loss. $s2$ also represents treatment. The treatment is the injection of IL-2 in the system.

Table~\ref{Tab:GillespieCase2} shows the mathematical model converted into reactions for the Gillespie algorithm model.
The first column of the table displays the original mathematical equation, followed by the equivalent reactions and rate laws in the subsequent columns.

\begin{figure}[!htpb]
  \resizebox{16cm}{!}{\includegraphics{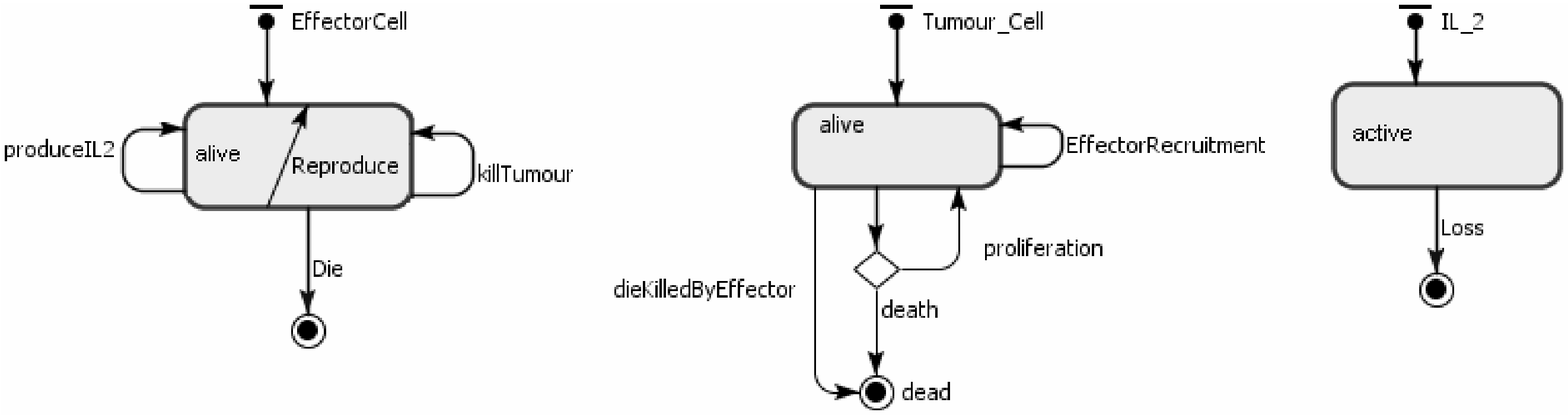}}
 \caption{{\bf ABMS state charts for the agents of case 2}}
 \label{fig:ABSModelThreeEquation}
\end{figure}

As described in~\cite{Figueredo2013a}, the agents represent the effector cells, tumour cells and IL-2. Their behaviours are shown in Table~\ref{Tab:ABSModelThreeEquation}. The state charts for each agent type are shown in Figure~\ref{fig:ABSModelThreeEquation}. The ABMS model rates are the same as those defined in the mathematical model and are given in Table~\ref{Tab:TumourEffectorTransitionRatesCalculations}. In the transition rate calculations, the variable $TotalTumour$ corresponds to the total number of tumour cell agents, the variable $TotalEffector$ is the total number of effector cell agents and $TotalIL\_2$ is the total number of IL-2 agents. In the simulation model, apart from the agents, there are also two events: the first event adds effector cell agents according to the parameter $s1$ and the second one adds IL-2 agents according to the parameter $s2$.

\subsubsection*{Experimental Design for the Simulation}

The experiment is conducted assuming the same parameters as those of the mathematical model (Table~\ref{Tab:CaseStudy2:ParametersModelCase2}). For the ABMS and the Gillespie algorithm model, the simulation is run five hundred times and the average outcome value for these runs is collected. Each run simulates a period equivalent to six hundred days, following the same time span used for the numerical simulation of the mathematical model.

\subsubsection*{Results and Discussion}
\label{sec:ResultsCase2}

\begin{figure}[!ht]
  \centering
  \resizebox{16.5cm}{!}{\includegraphics{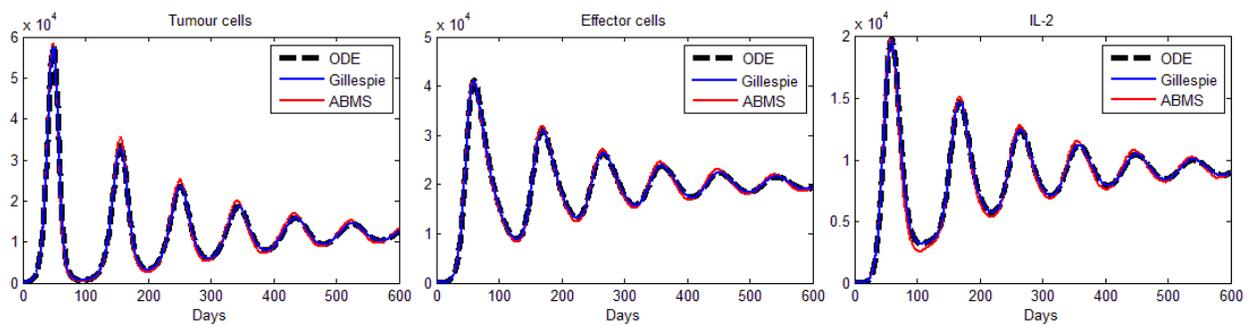}}
  \caption{{\bf Results for the second case study: tumour cells, effector cells and IL-2}}
  \label{fig:ResultsCase2}
\end{figure}

The results obtained are shown in Figure~\ref{fig:ResultsCase2} for tumour cells (left), effector cells (middle) and IL-2 (right), respectively. As the figure reveals, the results for all populations are analogous; the growth and decrease of all populations occur at similar times for all approaches. ABMS has a little more variability in the results, specially regarding IL-2. We believe that for this case, the large population sizes (around $10^4$) produce a lower variability in the outcomes of the stochastic approaches. For statistical comparison, results were contrasted by applying a non-linear mixed effect model, as shown next.

The local maxima sequence follows a second order polynomial function of the form:
\begin{equation}
f(t)=c+a(t-b)^2
\end{equation}

The local minima sequence follows a second order polynomial function of the form:
\begin{equation}
f(t)=c-a*(t-b)^2
\end{equation}

\begin{figure} [!ht]
\centering
  \resizebox{13cm}{!}{\includegraphics{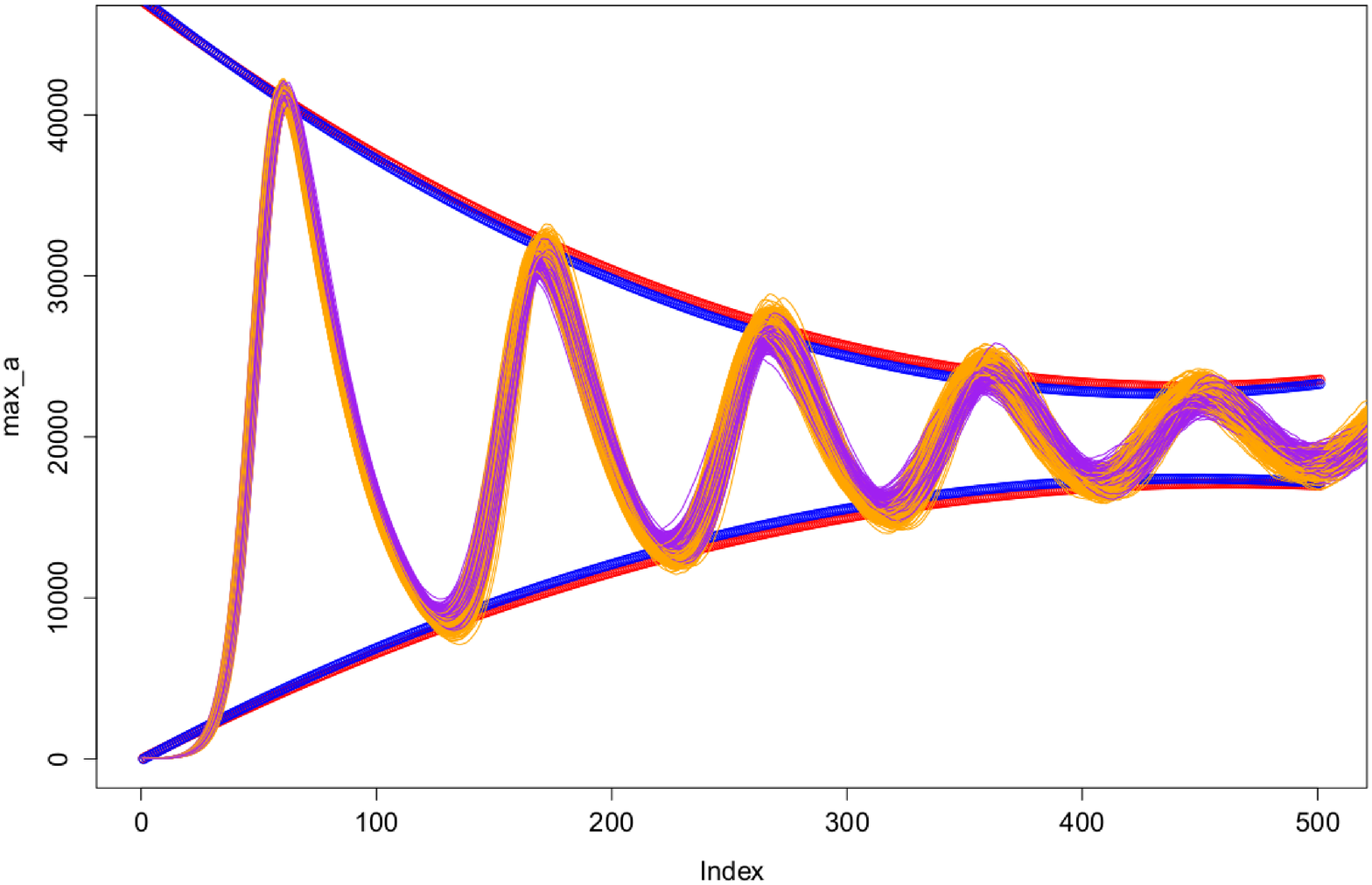}}
\caption{{\bf Illustration of the regression models fit for the sequences of the local maxima and local minima for the two different simulation techniques.  The Gillespie simulations are plotted in purple with the mixed effect models plotted in blue. The ABMS simulation runs are plotted in orange with the mixed effect models plotted in red.}}

\label{cs2_f}
\end{figure}

For the mixed-effect model, we consider $a$ and $b$ to have fixed effects based on the type of simulation (e.g., ABMS or Gillespie algorithm) and $a$ and $b$ to have random effects based on the individual simulation run.  The results of the mixed effect model are presented in Tables \ref{cs_2a} and \ref{cs_2b}. It can been seen that there is a significant difference between the $a$ and $b$ parameter values for the two different techniques.  We therefore reject the null hypotheses and accept that there is a significance difference between the two techniques in terms of the sequence of maxima and sequence of minima, at a 1\% significance level. Furthermore, the results show that the ABMS simulations tend to have larger local maxima and smaller local minima, which is clear in Figure \ref{cs2_f}.

\subsubsection*{Summary}

Interactions between tumour cells, effector cells and the cytokine IL-2 were considered to investigate the potential differences and similarities of ABMS and Gillespie algorithm outcomes. Statistical comparison between the Gillespie and the ABMS results show a significant difference in the outcomes. Compared to the original ODE model used as validation, ABMS displayed a little more variability in the results, whereas the Gillespie algorithm followed mostly the same patterns as those produced by the ODEs for all populations in the simulation. As for these simulations a bigger number of individuals was required, it was also observed that, regarding the use of computational resources, ABMS was far more time- and memory-consuming than the Gillespie approach.

\subsection*{Case 3: Interactions between Tumour Cells, Effector Cells, IL-2 and TGF-$\beta$}
\label{Case3}

Case study three comprises interactions between tumour cells and immune effector cells, as well as the immune-stimulatory and suppressive cytokines IL-2 and TGF-$\beta$~\cite{Arciero:2004}. According to the ODE model developed by Arciero {\it et al.} in~\cite{Arciero:2004}, TGF-$\beta$ stimulates tumour growth and suppresses the immune system by inhibiting the activation of effector cells and reducing tumour antigen expression.

The mathematical model is described by the differential equations below:

\begin{equation}
\label{Eq:cap:CaseStudy2:Case3:1}
 \frac{dE}{dt} = \frac{cT}{1 + \gamma S} - \mu_1E + \left ( \frac{p_1EI}{g_1+I} \right ) \left ( p_1 - \frac{q_1S}{q_2 + S} \right )
\end{equation}

Equation~\ref{Eq:cap:CaseStudy2:Case3:1} describes the rate of change for the effector cell population E. According to~\cite{Arciero:2004}, {\it effector cells are assumed to be recruited to a tumour site as a direct result of the presence of tumour cells}. The parameter c in $\frac{cT}{1 + \gamma S}$ represents the antigenicity of the tumour, which measures the ability of the immune system to recognize tumour cells. The presence of TGF-$\beta$ ($S$) reduces antigen expression, thereby limiting the level of recruitment, measured by the inhibitory parameter $\gamma$. The term $\mu_1E$ represents loss of effector cells due to cell death. The proliferation term $\left ( \frac{p_1EI}{g_1+I} \right ) \left ( p_1 - \frac{q_1S}{q_2 + S} \right )$ asserts that effector cell proliferation depends on the presence of the cytokine IL-2 and is decreased
when the cytokine TGF-$\beta$ is present. $p_1$ is the maximum rate of effector cell proliferation in the absence of TGF-$\beta$, $g_1$ and $q_2$ are half-saturation constants, and $q_1$ is the maximum rate of anti-proliferative effect of TGF-$\beta$.

\begin{equation}
\label{Eq:cap:CaseStudy2:Case3:2}
 \frac{dT}{dt} = aT \left ( 1 - \frac{T}{K} \right ) - \frac{a_aET}{g_2 + T} + \frac{p_2ST}{g_3+S}
\end{equation}

Equation~\ref{Eq:cap:CaseStudy2:Case3:2} describes the dynamics of the tumour cell population. The term $aT \left ( 1 - \frac{T}{K}\right )$
represents a logistic growth dynamics with intrinsic growth rate $a$ and carrying capacity $K$ in the absence of effector cells and TGF-$\beta$. The term $\frac{a_aET}{g_2 + T}$ is the number of tumour cells killed by effector cells. The parameter $a_a$ measures the strength of the immune response to tumour cells. The third term $\frac{p_2ST}{g_3+S}$ accounts for the increased growth of tumour cells in the presence of
TGF-$\beta$. $p_2$ is the maximum rate of increased proliferation and $g_3$ is the half-saturation constant, which indicates a limited response of tumour cells to this growth-stimulatory cytokine~\cite{Arciero:2004}.

\begin{equation}
\label{Eq:cap:CaseStudy2:Case3:3}
 \frac{dI}{dt} = \frac{p_3ET}{(g_4 + T)(1+ \alpha S)} - \mu_2I
\end{equation}

The kinetics of IL-2 are described in equation~\ref{Eq:cap:CaseStudy2:Case3:3}. The first term $\frac{p_3ET}{(g_4 + T)(1+ \alpha S)}$ represents IL-2 production which reaches a maximal rate of $p_3$ in the presence of effector cells stimulated by their interaction with the tumour cells. In the absence of TGF-$\beta$, this is a self-limiting process with half-saturation constant $g_4$~\cite{Arciero:2004}. The presence of TGF-$\beta$ inhibits IL-2 production, where the parameter $\alpha$ is a measure of inhibition. Finally, $\mu_2I$ represents the loss of IL-2.

\begin{equation}
\label{Eq:cap:CaseStudy2:Case3:4}
 \frac{dS}{dt} = \frac{p_4T^2}{\theta^2+T^2} - \mu_3S
\end{equation}

Equation~\ref{Eq:cap:CaseStudy2:Case3:4} describes the rate of change of the suppressor cytokine, TGF-$\beta$. According to~\cite{Arciero:2004}, {\it experimental evidence suggests that TGF-$\beta$ is produced in very small amounts when tumours are small enough to receive ample nutrient from the surrounding tissue. However, as the tumour population grows sufficiently large, tumour cells suffer from a lack of oxygen and begin to produce TGF-$\beta$ in order to stimulate angiogenesis and to evade the immune response once tumour growth resumes}. This switch in TGF-$\beta$ production is modelled by the term $\frac{p_4T^2}{\theta^2+T^2}$, where $p_4$ is the maximum rate of TGF-$\beta$ production and $\tau$ is the critical tumour cell population in which the switch occurs. The decay rate of TGF-$\beta$ is represented by the term $\mu_3S$.

Table~\ref{Tab:GillespieCase3} presents the Gillespie algorithm model used for our simulations. The model was obtained by converting the ODEs into reaction equations.

\begin{figure}[!hb]
\centering
  \resizebox{16cm}{!}{\includegraphics{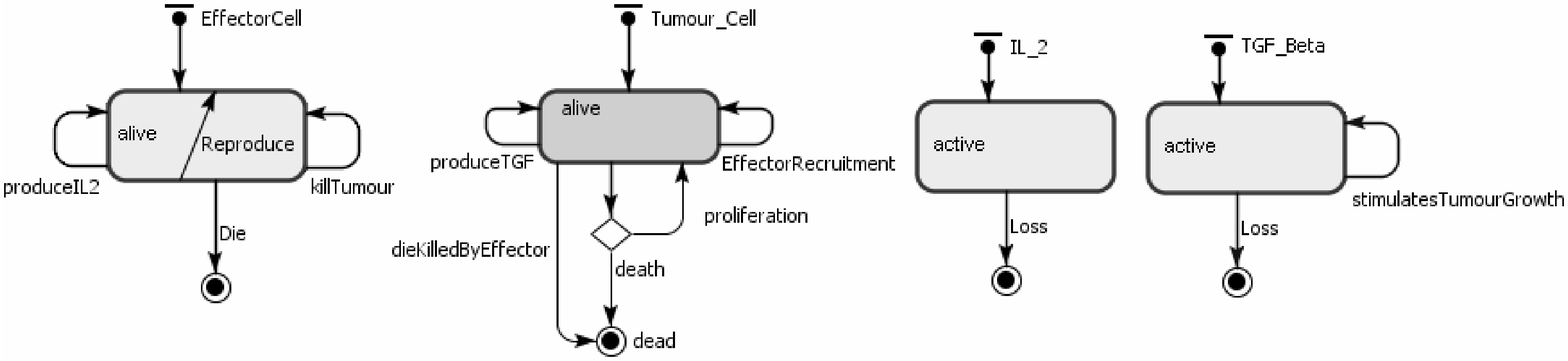}}
\caption{{\bf ABMS state charts for the agents of case 3}}
\label{fig:ABSModelFourEquation}
\end{figure}

The agents established for the ABMS represent the effector cells, tumour cells, IL-2 and TGF-$\beta$ populations, as described in~\cite{Figueredo2013a}. The agents' behaviour is defined in Table~\ref{Tab:ABSModelFourEquation}. The state charts for each agent type are illustrated in Figure~\ref{fig:ABSModelFourEquation}.

The ABMS model rates corresponding to the mathematical model are given in Table~\ref{Tab:TumourEffector4TransitionRatesCalculations}. In the transition rate calculations, the variable $TotalTumour$ corresponds to the total number of tumour cell agents; the variable $TotalEffector$ is the total number of effector cell agents, $TotalIL\_2$ is the total number of IL-2 agents and $TotalTGFBeta$ is the total TGF-$\beta$ agents. This model does not include events.

\subsection*{Experimental design for the simulation}

The experiment is conducted assuming the same parameters as those defined for the mathematical model (Table~\ref{Tab:CaseStudy2:ParametersModelCase3}). Similarly to the previous case studies, for the ABMS and Gillespie models the simulation is run five hundred times and the average outcome value for these runs is displayed as result. Each run simulates a period equivalent to six hundred days, following the time interval used for the numerical simulation of the mathematical model. The parameters used for the simulations of all approaches are shown in Table~\ref{Tab:CaseStudy2:ParametersModelCase3}.

\subsubsection*{Results and discussion}
\label{sec:ResultsCase3}

\begin{figure}[!ht]
  \centering
  \resizebox{16.5cm}{!}{\includegraphics{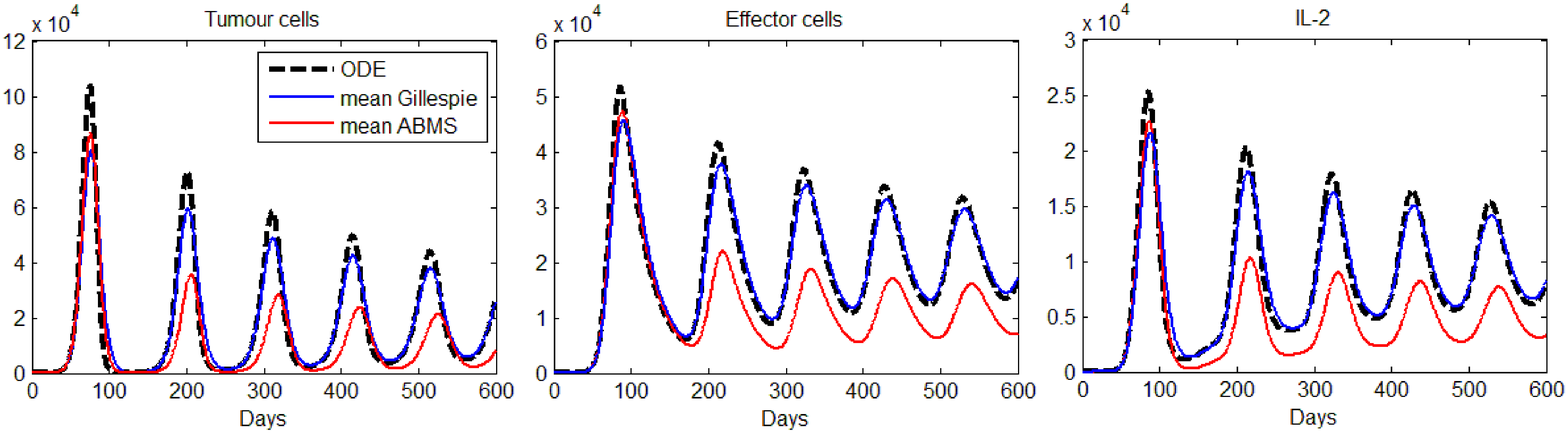}}
  \caption{{\bf Results for the third case study: mean values of tumour cells, effector cells and IL-2}}
  \label{fig:ResultsCase3MeanValues}
\end{figure}

\begin{figure}[!htpb]
  \centering
  \resizebox{13.8cm}{!}{\includegraphics{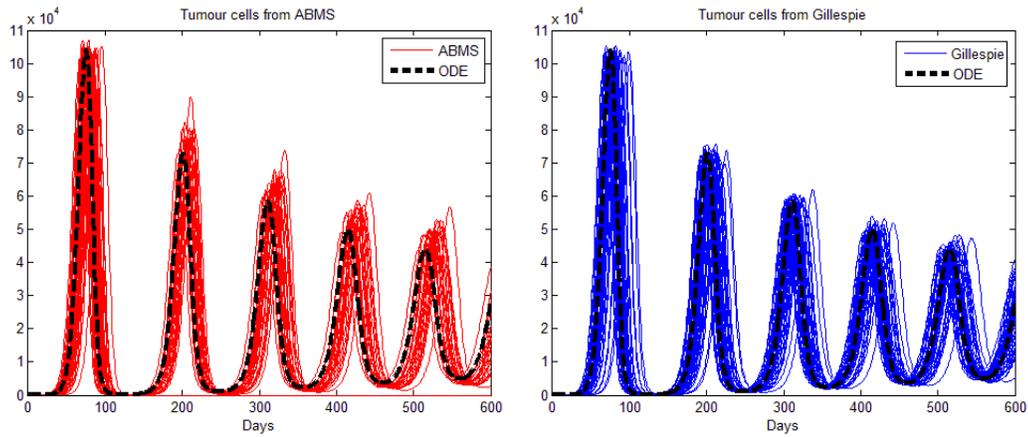}}
  \caption{{\bf Results for the third case study: 50 runs for tumour cells}}
  \label{fig:ResultsCase3Tumour50Runs}
\end{figure}

\begin{figure}[!htpb]
  \centering
  \resizebox{13.8cm}{!}{\includegraphics{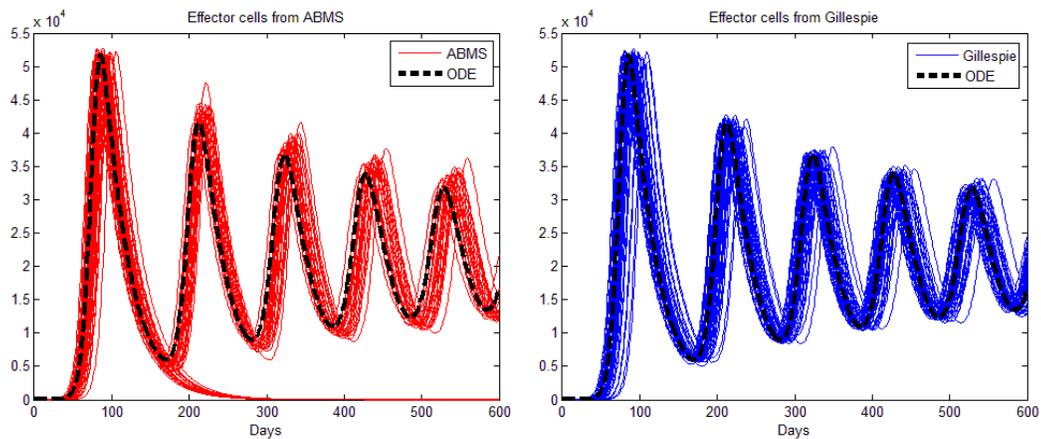}}
  \caption{{\bf Results for the third case study: 50 runs for effector cells}}
  \label{fig:ResultsCase3Effector50Runs}
\end{figure}

\begin{figure}[!ht]
  \centering
  \resizebox{13.8cm}{!}{\includegraphics{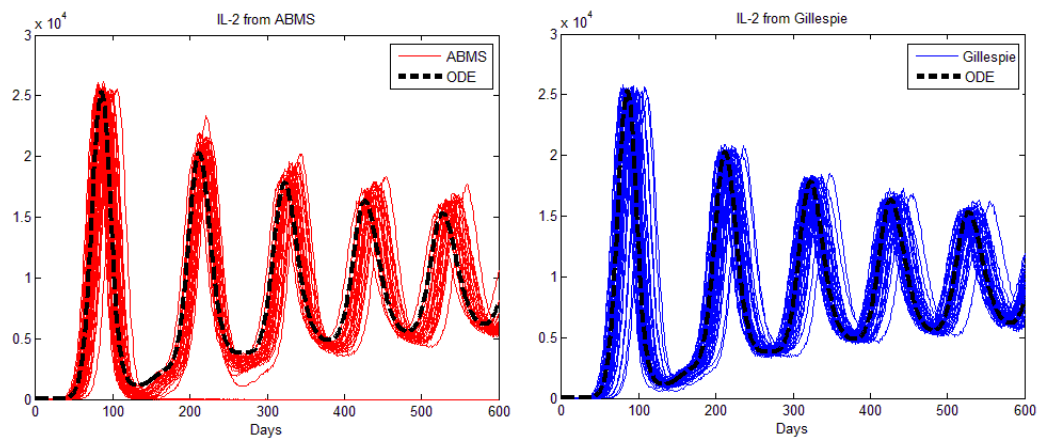}}
  \caption{{\bf Results for the third case study: 50 runs for IL2 cells}}
  \label{fig:ResultsCase3IL250Runs}
\end{figure}

\begin{figure}[!ht]
  \centering
  \resizebox{16cm}{!}{\includegraphics{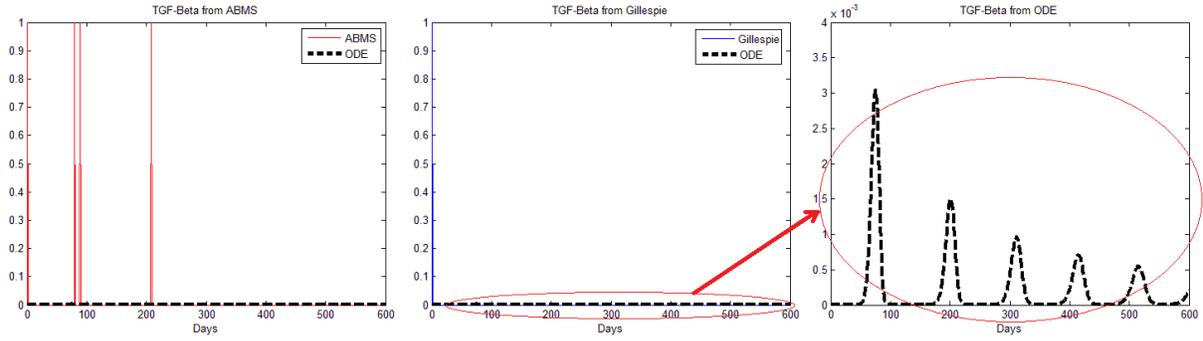}}
  \caption{{\bf Results for the third case study: 50 runs for TGF-$\beta$. The figure on the right is a zoomed version of the figure in the centre.}}
  \label{fig:ResultsCase3TGF50Runs}
\end{figure}

The mean results of 500 runs for the Gillespie algorithm and the ABMS contrasted with the ODE model are shown in Figure~\ref{fig:ResultsCase3MeanValues}. The left graph in the figure presents the outcomes for tumour cells; the graph in the middle shows the outputs for effector cells; the graph on the right shows the mean IL-2 outcomes (the TGF-$\beta$ results have some particularities and therefore are discussed next). The figure shows that both Gillespie and ABMS do not match properly the original results from the mathematical model. Additionally, ABMS is far more dissimilar than what was anticipated. In order to understand why the mean values were that much different from what was expected, we plotted fifty individual runs for each approach, as shown in figures~\ref{fig:ResultsCase3Tumour50Runs}, ~\ref{fig:ResultsCase3Effector50Runs}, ~\ref{fig:ResultsCase3IL250Runs} and~\ref{fig:ResultsCase3TGF50Runs}. These runs illustrate the variations observed in both ABMS (left side of the figures) and Gillespie (right side of the figures) approaches, due to its stochastic character. In the figures, the ODE model results were also plotted (dashed black line) in order to highlight the range of variation produced by the stochastic approaches. As it can be observed in the figures, both Gillespie and ABMS outcomes produce various slightly distinct starting times for the growth of populations. In addition to these variations, for a few runs the populations in ABMS decreased to zero, as previously reported in~\cite{Figueredo2013a}. This behaviour was not reflected in the Gillespie algorithm results. This indicates that it is not always possible to replicate similar results within both approaches.

The use of ABS modelling has therefore led to the discovery of additional ``rare'' patterns, which we would have not been able to derive by using analytical methods or the dynamic Monte Carlo method, i.e. the Gillespie algorithm. These ``extreme cases'' found by ABMS suggest that there might be circumstances where the tumour cells are completely eliminated by the immune system, without the need of any cancer therapies.

We believe that ABMS when compared to Gillespie produces extra patterns because of the agents individual behaviour and their interactions. While ODEs and the Gillespie algorithm always use the same values for the parameters over the entire population aggregate, ABMS rates vary with time and number of individuals. Each agent is likely to have distinct numbers for their probabilities and therefore have its own memory of past events (Gilespie, however, does not encompass individual memory for its elements). The agents individual interactions, which give raise to the overall behaviour of the system, are also influenced by the scenario determined by the random numbers used. By running the ABMS multiple times with different sets of random numbers, the outcomes vary according to these sets and the emerging interactions of the agents also produce the rare outcome patterns.

For further statistical comparison of the results that follow the same pattern of behaviour for ABMS and Gillespie, a mixed-effect model is used. In 236 of the 500 ABMS simulations the tumour cell population dies out early in time.  The remaining 264 ABMS simulations (where the tumour cell population does not die out over the [0,600] time period) is compared with the 500 Gillespie simulations by a non-linear mixed effect model.

The local maxima sequence follows a second order polynomial function of the form:
\begin{equation}
f(t)=40311+a(t-b)^2
\end{equation}

The local minima sequence follows a second order polynomial function of the form:
\begin{equation}
f(t)=a*(t-b)^2
\end{equation}

For the mixed-effect model we considered  $a$ and $b$ to have fixed effects based on the type of simulation (e.g., ABMS or Gillespie algorithm) and $a$ and $b$ to have random effects based on the individual simulation run.  The results of the mixed effect model are presented in Tables \ref{cs_3a} and \ref{cs_3b}. It can been seen that there is a significant difference between the $a$ and $b$ parameter values for the two different techniques. We therefore reject the null hypotheses and accept that there is a significance difference between the two techniques in terms of the sequence of maxima and sequence of minima, at a 1\% significance level.

\begin{figure}[!ht]
\centering
\resizebox{12cm}{!}{\includegraphics{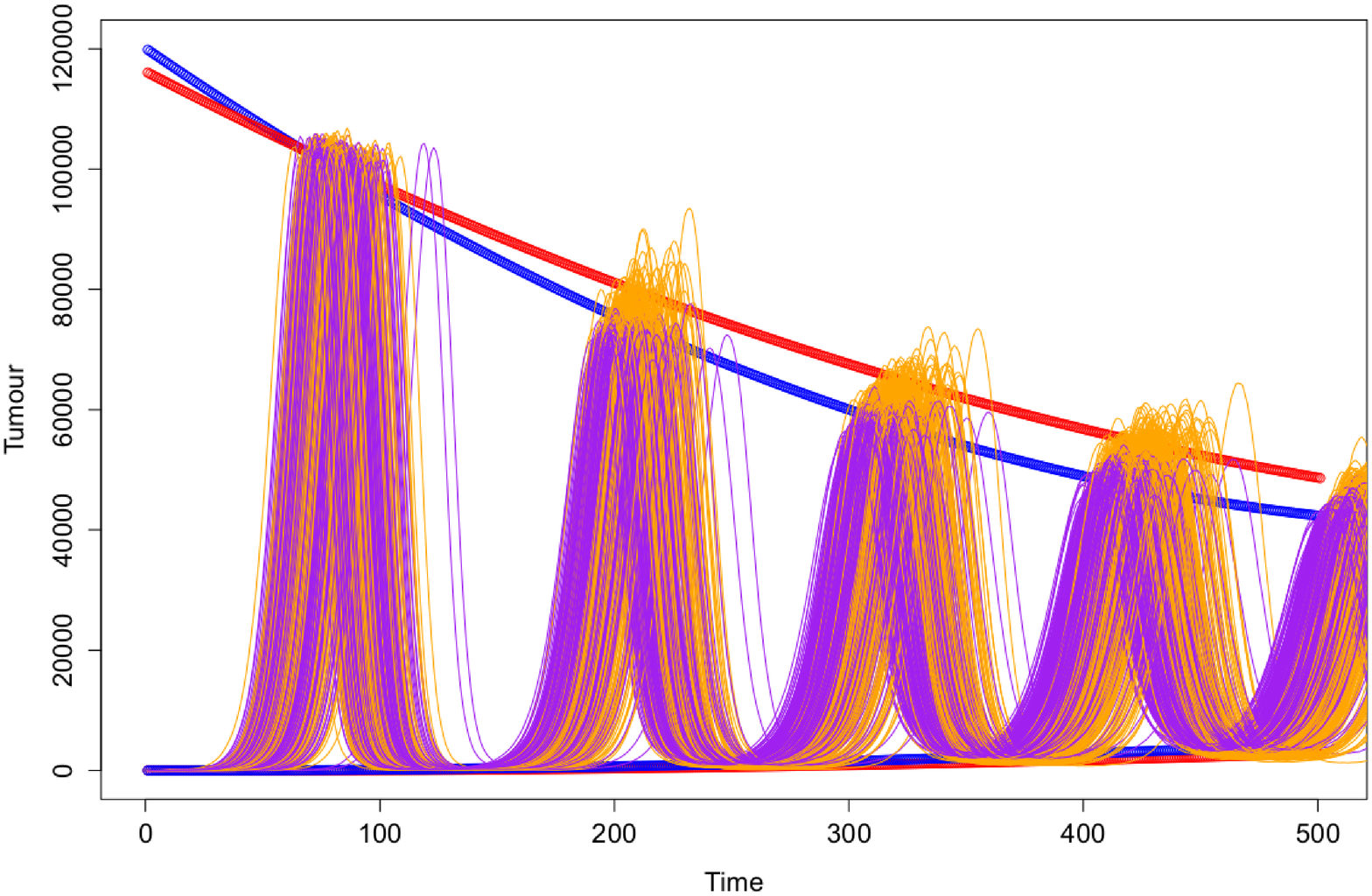}}
\caption{{\bf Illustration of the regression models fit for the sequences of the local maxima and local minima for the two different simulation techniques. The Gillespie simulations are plotted in purple with the mixed effect models plotted in blue. The ABMS simulation runs are plotted in orange with the mixed effect models plotted in red.}}
\label{cs3_f}
\end{figure}

Tables \ref{cs_3a}---\ref{cs_3b} and Figure \ref{cs3_f} show that the sequence of local maxima of the ABMS diverge from that of the Gillespie algorithm over time. In the ABMS the tumour cells tend to increase to a larger count than the Gillespie algorithm simulations causing the function of local maxima for the ABMS to be significantly greater than the Gillespie.  A possible explanation for this, as mentioned previously, is the fact that agents have memory and therefore the rates (death, proliferation, etc) for a certain cell are determined by the cells (and their proportions) present in the system at the moment the cell was created. For the Gillespie algorithm, instead, the rates are applied globally to the entire population and remain constant over the simulation course. Consequently, for Gillespie, the individuals do not keep a record of the previous population dynamics. This explanation is supported by the observation that the function describing the sequence of local minima of the ABMS is significantly lower than the Gillespie algorithm over time, as the same argument would account for the ABMS simulations reaching lower levels.

Regarding the  TGF-$\beta$ outcomes, the ODEs results reveal numbers smaller than one (Figure~\ref{fig:ResultsCase3TGF50Runs} on the right), which is not possible to achieve with the ABMS and the Gillespie algorithm. The simulation results regarding these molecules are therefore completely different for both stochastic approaches. By observing the multiple runs graph of ABMS, however, results indicate that the TGF-$\beta$ grows at around 100 and 200 days, which resembles what occurs in the ODE simulation for the first two peaks of TGF-$\beta$ concentration. This suggests that ABMS, as opposite to Gillespie, is capable of capturing some of the behaviours of the analytical results even when the outputs are different. We believe that these observations need to be further investigated in order to determine whether this happens in other case studies. In addition, it is necessary to investigate in what circumstances and range of values ABMS is still capable to reflect behaviours of numbers smaller than one agent present in the ODE model.

\subsection*{Summary}

The third case study simulations investigated interactions between effector cells, tumour cells and two types of cytokines, namely IL-2 and TGF-$\beta$. When compared to the original ODE results, both Gillespie algorithm and ABMS produced more variability in the outcomes. For each of the five hundred runs, a slightly different start of population growth was observed. In addition, ABMS produced extra patterns not observed in the original mathematical model and in the Gillespie results. These extra patterns have been reported previously in~\cite{Figueredo2013a} and with the present work we wanted to find out whether the Gillespie algorithm simulation results would be as informative. This indicates that, for this case study, both methods should not be employed interchangeably, as some extra possible population patterns of behaviour might not be uncovered without ABMS. We believe that these emergent examples occur due to the individual interactions of the agents and their chaotic character. With these results, we answer our third research question that it is not possible in this case to obtain extreme patterns using the Gillespie algorithm.

\section*{Conclusions}
\label{Conclusions}

In this work, we employed three case studies to investigate circumstances where we can use ABMS and the Gillespie algorithm interchangeably. We aimed at reproducing the variability embedded in the ABMS systems to the mathematical formulation and verify whether results resemble. Current literature regarding the comparison of the Gillespie algorithm and ABMS is scarce and we wanted therefore to answer the questions: (1) Does the Gillespie algorithm produce similar results to ABMS? (2) Can these two methods be used interchangeably for our case studies? (3) Does the Gillespie algorithm also find the extra patterns revealed by the ABMS in the third case study? The case studies investigated regarded models with different characteristics, such as population sizes, modelling effort demanded and model complexity.

The first case study involved interactions with general immune effector cells and tumour cells. Four different scenarios regarding distinct sets of parameters were investigated and in the first three scenarios treatment was included.
ABMS and Gillespie produced different results for all scenarios. It appears that two major characteristics of this model influenced the differences obtained: (1) The small quantities of individuals considered in the simulations (especially regarding the effector population size, which was always smaller than ten) that significantly increased the variability of both stochastic approaches; and (2) the stochasticity of the Gillespie algorithm is applied to the aggregates, while in the ABMS there is individual variability.

Case study 2 referred to the investigation of a scenario containing interactions between effector cells, cytokines IL-2 and tumour cells. For this case the Gillespie and ABMS approaches produced similar outcome curves, which also matched the pattern of behaviour of the mathematical model used for validation. As populations’ sizes  had a magnitude of $104$ individuals, the erratic behaviour of both stochastic approaches was no longer evident in the outcomes. However, although results seemed similar, further statistical tests reject their similarity hypothesis. It was observed that, for case 2 in general, ABMS simulation outcome curves tend to have larger local maxima and smaller local minima.

Case study 3 includes the influence of the cytokine TGF-$\beta$ in the interactions between effector cells, cytokines IL-2 and tumour cells from the previous case. The simulation outcomes for the ABMS were mostly following the same pattern as those produced by the Gillespie algorithm, although the results were statistically different. In addition, Gillespie failed to replicate the alternative outcomes found by the ABMS. This indicates that for this case study the ABMS results are more informative, as they illustrate another set of possible dynamics to be validated in real-world. Furthermore, ABMS was also able to indicate two peaks where TGF-$\beta$ concentrations have grown, although the corresponding values in the mathematical model were smaller than one.

In response to our research questions, we conclude that regarding the interchangeable use of Gillespie and ABMS, population size has a positive impact on result similarity. This means that bigger populations tend to result in close simulation output patterns. However, the stochasticity of both approaches and the memory present in the ABMS produce outcome differences which are statistically significant, although visually the outcomes look similar. Finally, the emergent behaviour of ABMS can contribute additional insight (extra patterns), which was not obtained by the aggregate stochasticity present in Gillespie, given its incapacity of retaining memory of past events for their elements.

\section*{Acknowledgments}
This work was supported by the Advanced Data Analysis Centre (ADAC) at the University of Nottingham.

\bibliography{tese}

\newpage
\section*{Figure Legends}

\clearpage

\section*{Tables}

\begin{table}[!htpb]
\centering
\caption{{\bf Case studies considered}}
\begin{tabular}{|l|c|c|c|}
\hline
Case Study                             & Number of populations             & Population size  & Complexity \\
\hline
\hline
1) Tumour/Effector                     & 2                                 &  5 to 600        &  Low       \\
\hline
2) Tumour/Efector/IL-2                 & 3                                 & $10^4$           &  Medium    \\
\hline
3) Tumour/Effector/IL-2/ TGF-$\beta$   & 4                                 &  $10^4$          &  High       \\
\hline
\end{tabular}
\label{Tab:CaseStudies}
\end{table}

\begin{table}[!htpb]
\centering
\caption{{\bf Reactions for case 1}}
\begin{tabular}{|l|l|l|}
\hline
Phenomenon                           & Reaction equation             & Rate law (per cell) \\
\hline
\hline
Tumour cell birth                    & $T \rightarrow 2 \times T$    &  $aT$                \\
\hline
Tumour cell death                    & $2 * T \rightarrow T$         &  $abT^{2}$            \\
\hline
Tumour cell death by effector cells  & $T + E \rightarrow E$         &  $nTE$              \\
\hline
Effector proliferation               & $E \rightarrow 2 \times E; T$  &  $\frac{pTE}{g+T}$  \\
\hline
Effector death by fighting tumour    & $T + E \rightarrow T$          &  $mTE$              \\
\hline
Effector death                       & $E \rightarrow$                &  $dE$               \\
\hline
Effector supply                      & $\rightarrow E$                &  $s$                 \\
\hline
\end{tabular}
\label{Tab:GillespieCase1}
\end{table}

\begin{table}[!htpb]
\begin{center}
\caption{{\bf Agents' parameters and behaviours for case 1}}
\begin{tabular}{|l|l|l|l|}
\hline
Agent          & Parameters              &  Reactive behaviour       & Proactive behaviour                \\
\hline \hline
\multirow{4}{*}
{Tumour Cell}  & $a$ and $b$             & Dies (with age)                     &  \\
               \cline{2-4}
               & $a$ and $b$             &                           & Proliferates              \\
               \cline{2-4}
               & $m$                     &                           & Damages effector cells\\
               \cline{2-4}
               & $n$                     &  Dies killed by effector cells &                 \\
\hline
\hline
\multirow{4}{*}
{Effector Cell}& $m$                     &  Dies (with age)            & \\
               \cline{2-4}
               & $d$                     &  Dies per apoptosis            & \\
               \cline{2-4}
               & $p$ and $g$             &                           & Proliferates             \\
               \cline{2-4}
               & $s$                     &  Is injected as treatment &\\
\hline
\end{tabular}
\label{Tab:Case1:AgentsBehaviours}
\end{center}
\end{table}

\begin{table}[!htpb]
\begin{center}
\caption{{\bf Transition rates calculations from the mathematical equations for case 1}}
\begin{tabular}{|l|l|l|l|}
\hline
Agent                &  Transition              & Mathematical equation                               & Transition rate       \\
\hline \hline
\multirow{4}{*}
{Tumour Cell}       &  proliferation           & $aT(1 - Tb)$                                        & $a - (TotalTumour.b)$   \\

                     \cline{2-4}

                     &  death                   & $aT(1 - Tb)$                                        & $a - (TotalTumour.b)$   \\

                     \cline{2-4}

                     &  dieKilledByEffector    & $nTE$                                               & $n.TotalEffectorCells$\\

                     \cline{2-4}

                     &  causeEffectorDamage     & $mTE$                                               & $m$\\

\hline
\hline

\multirow{3}{*}
{Effector Cell}      & Reproduce            & $\frac{pTE}{g+T}$ & $\frac{p. TotalTumourCells}{g+TotalTumourCells}$\\

                     \cline{2-4}

                     & DiePerAge               & $dE$                      & $d$\\

                     \cline{2-4}
                     & DiePerApoptosis          & $mTE$                     & message from tumour   \\
\hline
\end{tabular}
\label{Tab:TumourEffectorTransitionRatesCalculations1}
\end{center}
\end{table}

\begin{table}[!htpb]
\centering
\caption{{\bf Simulation parameters for the four different scenarios under investigation. For the other parameters, the values are the same in all experiments, i.e. $a=1.636$, $g=20.19$, $m=0.00311$, $n=1$ and $p=1.131$.}}
\begin{tabular}{|l|c|c|c|}
\hline
Scenario  & b     &  d     & s      \\
\hline \hline
1         &0.002 & 0.1908  & 0.318  \\
\hline
2         &0.004 & 2       & 0.318  \\
\hline
3         &0.002 & 0.3743  & 0.1181 \\
\hline
4         &0.002 & 0.3743  & 0  \\
\hline
\end{tabular}
\label{Tab:CaseStudy1:TumourEffectorsParameters}
\end{table}

\begin{table} \centering
\caption{{\bf The fixed parameter values returned by the mixed-effect model and their significance.}}
\label{tab:statsCase1}
\begin{tabular}{|l|l|l|l|l|}
\hline
Method & Parameter & Value & std error & p-value \\
\hline
\hline
ABS & $a$ & 0.03393 & 0.0004397 &     0\\
\hline
SODE & $a$ & 0.02925&0.0006126 &      0  \\
\hline
ABS & $b$ & 41.15847& 0.6149638 &    0\\
\hline
SODE & $b$ &52.07957 & 1.0328222 &       0 \\
\hline
\end{tabular}
\end{table}

\begin{table}[!htpb]
\centering
\caption{{\bf Reactions for case 2}}
\begin{tabular}{|l|l|l|}
\hline
Name                                 & Reaction equation       & Rate law (per cell) \\
\hline
\hline
Effector cell recruitment            & $\rightarrow E;  T$           &  $cT$                      \\
\hline
Effector cell proliferation          & $E \rightarrow 2 \times E; I$ & $\frac{p_{1}E.I_L}{(g_1+I_L)}$\\
\hline
Effector cell death                  & $E \rightarrow $              &  $\mu_2E$    \\
\hline
Effector cell supply                 & $\rightarrow E$              &  $s1$        \\
\hline
Tumour cell birth                    & $T \rightarrow 2 \times T$    &  $aT$        \\
\hline
Tumour cell death                    & $2 * T \rightarrow T$         &  $abT^{2}$ \\
\hline
Tumour cell death by effector cells  & $T + E \rightarrow E; T$      &$\frac{a_aTE}{g_2+T}$\\
\hline
IL-2 production                      & $\rightarrow I;  E T$         & $\frac{p_{2}ET}{g_3+T}$ \\
\hline
IL-2 decay                           & $I \rightarrow$               &  $\mu_3$          \\
\hline
IL-2 supply                          & $\rightarrow I$               &  $s2$  \\
\hline
\end{tabular}
\label{Tab:GillespieCase2}
\end{table}

\begin{table}[!ht]
\centering
\caption{{\bf Agents' parameters and behaviours for case 2}}
\begin{tabular}{|l|l|l|l|}
\hline
Agent          & Parameters             & Reactive behaviour            &  Proactive behaviour            \\
\hline \hline
\multirow{6}{*}
{Effector Cell}& $mu2$                  & Dies                       & \\
               \cline{2-4}
               & $p1$ and $g1$          &                            & Reproduces                      \\
               \cline{2-4}
               & $c$                    & Is recruited               &                                 \\
               \cline{2-4}
               & $s1$                   & Is injected as treatment   &                                 \\
               \cline{2-4}
               & $p2$ and $g3$          &                            & Produces IL-2\\
               \cline{2-4}
               & $aa$ and $g2$          &                            & Kills tumour cells\\

\hline
\hline
\multirow{4}{*}
{Tumour Cell}  & $a$ and $b$            &  Dies  & \\
               \cline{2-4}
               & $a$ and $b$            &                             & Proliferates         \\
               \cline{2-4}
               & $aa$ and $g2$          &  Dies killed by effector cells&                      \\
               \cline{2-4}
               & $c$                    &                               & Induces effector recruitment\\
\hline
\hline
\multirow{3}{*}
{IL-2}         &  $p2$ and $g3$         &  Is produced   &\\
               \cline{2-4}
               &  $mu3$                 &  Is lost       &\\
               \cline{2-4}
               &  $s2$                  &  Is injected   &\\

\hline
\end{tabular}
\label{Tab:ABSModelThreeEquation}
\end{table}

\begin{table}[!htpb]
\centering
\caption{{\bf Transition rates calculations from the mathematical equations for case 2}}
\begin{tabular}{|l|l|l|l|}
\hline
Agent                &  Transition              & Mathematical equation             & Transition rate       \\
\hline \hline
\multirow{4}{*}
{Effector Cell}
                     & Reproduce                & $\frac{p_1.I_LE}{g1+IL\_2}$        & $\frac{p_1.TotalIL\_2.TotalEffector}{g1+TotalIL\_2}$\\

                     \cline{2-4}

                     & Die                      & $\mu_2E$                           & $mu2$\\

                     \cline{2-4}

                     & killTumour               & $\frac{a_aET}{g2 + T}$             & $aa \frac{TotalTumour}{g2 + TotalTumour}$\\

                     \cline{2-4}

                     & ProduceIL2               &$\frac{p2ET}{g3 + T}$               & $\frac{p2.TotalTumour}{g3 + TotalTumour}$\\

\hline
\hline

\multirow{3}{*}
{Tumour Cell}         &  Reproduce               & $aT(1 - bT)$                        & $a - (TotalTumour.b)$   \\

                     \cline{2-4}

                     &  Die                     & $aT(1 - bT)$                        & $a - (TotalTumour.b)$ \\

                     \cline{2-4}

                     &  DieKilledByEffector     & $\frac{a_aTE}{g2+T}$                & message from effector\\

\hline
\hline

IL-2                & Loss                      & $\mu_3I_L$                             & $mu3$\\
\hline
\end{tabular}
\label{Tab:TumourEffectorTransitionRatesCalculations}
\end{table}

\begin{table}[ht!]
\centering
\caption{{\bf Parameter values for case 2}}
\begin{tabular}{|l|l|}
\hline
Parameter & Value\\
\hline
\hline
a         & 0.18\\
\hline
b & 0.000000001\\
\hline
c & 0.05\\
\hline
aa & 1\\
\hline
g2 & 100000\\
\hline
s1 & 0\\
\hline
s2 & 0 \\
\hline
mu2 & 0.03\\
\hline
p1 & 0.1245\\
\hline
g1 & 20000000\\
\hline
p2 & 5\\
\hline
g3 & 1000\\
\hline
mu3 & 10\\
\hline
\end{tabular}
 \label{Tab:CaseStudy2:ParametersModelCase2}
\end{table}

\begin{table}[!htpb]
\centering
\caption{{\bf The results of the mixed model for the sequence of local maxima in case study 2}}
\label{cs_2a}
\begin{tabular}{|l|l|l|l|l|}
\hline
Technique & Parameter & Value & Std Error & p-value \\
\hline
\hline
ABMS & a & 0.122 & 0.00073 & 0 \\
\hline
Gillespie & a &  0.131 & 0.00102 & 0 \\
\hline
ABMS & b & 442.249 & 1.160664 & 0 \\
\hline
Gillespie & b &  432.243 & 1.52861 & 0 \\
\hline
ABMS & c & 23149.344 & 23.67208 & 0 \\
\hline
Gillespie & c &  22694.685 & 31.88635 & 0 \\
\hline
\end{tabular}
\end{table}

\begin{table}[!htpb]
\centering
\caption{{\bf The results of the mixed model for the sequence of local minima in case study 2}}
\label{cs_2b}
\begin{tabular}{|l|l|l|l|l|}
\hline
Technique & Parameter & Value & Std Error & p-value \\
\hline
\hline
ABMS & a & 0.080 & 0.00033 & 0 \\
\hline
Gillespie & a &  0.088 & 0.00047 & 0 \\
\hline
ABMS & b & 462.004 & 1.12224 & 0 \\
\hline
Gillespie & b &  444.888 & 1.46963 & 0 \\
\hline
ABMS & c & 17118.133 & 25.43450 & 0 \\
\hline
Gillespie & c &  17416.363 & 34.44740 & 0 \\
\hline
\end{tabular}
\end{table}

\begin{table}[!htpb]
\centering
\caption{{\bf Reactions for case 3}}
\begin{tabular}{|l|l|l|}
\hline
Name                                 & Equation                        & Rate law (per cell)                     \\
\hline
\hline
Effector cell recruitment            & $  \rightarrow E;  T S $        & $\frac{c \times T}{1+\gamma \times S}$ \\
\hline
Effector death                       & $E \rightarrow$                 & $\mu_1 \times E$                       \\
\hline
Effector proliferation               & $T + E \rightarrow E;  T$       & $\left (\frac{p\times I \times E}{g+I}\right ) \times \left ( p - (\frac{q1 \times S}{q2+S} \right )$       \\
\hline
Tumour cell growth                   & $T \rightarrow 2 \times T$      & $a\times T$     \\
\hline
Tumour cell death                    & $2 \times T \rightarrow T$      & $\frac{aT^{2}}{K}$      \\
\hline
Tumour cell death by effector cells  & $T + E \rightarrow E;  T$       & $\frac{a_{a}TE}{g_{2}+T}$      \\
\hline
Tumour growth caused by TGF-$\beta$  & $T \rightarrow 2 \times T;  S$  & $\frac{p_2 \times S \times T}{g_3+S}$  \\
\hline
IL-2 production                      & $\rightarrow I;  E T S $        & $\frac{p_3 \times E \times T}{(g4+T)(1+ \alpha S)}$ \\
\hline
IL-2 decay                           & $I \rightarrow$                 & $\mu_2 \times I$\\
\hline
TGF-$\beta$ production               & $\rightarrow I;  T$             & $\frac{p_4T^2}{\theta^{2} + T^{2}}$\\
\hline
TGF-$\beta$ decay                    & $S \rightarrow$                 & $\mu_3 \times S$\\
\hline
\end{tabular}
\label{Tab:GillespieCase3}
\end{table}

\begin{table}[!ht]
\centering
\caption{{\bf Agents' parameters and behaviours for case 3}}
\begin{tabular}{|l|l|l|l|}
\hline
Agent          & Parameters                    & Reactive behaviour            &  Proactive behaviour            \\
\hline \hline
\multirow{4}{*}
{Effector Cell}& $mu1$                         & Dies                       & \\
               \cline{2-4}
               & $p1$, $g1$, $q1$ and $q2$     &                            & Reproduces                      \\
               \cline{2-4}
               & $c$                           & Is recruited               &                                 \\
               \cline{2-4}
               & $aa$ and $g2$                 &                            & Kills tumour cells\\

\hline
\hline
\multirow{6}{*}
{Tumour Cell}  & $a$                           &  Dies  & \\
               \cline{2-4}
               & $a$                           &                             & Proliferates         \\
               \cline{2-4}
               & $aa$ and $g2$                 &  Dies killed by effector cells&                      \\
               \cline{2-4}
               & $g3$ and $p2$                 &  Has growth stimulated       &\\
               \cline{2-4}
               & $p4$ and $tetha$              &                              & Produces TGF-$\beta$\\
               \cline{2-4}
               & $c$                           &                              & Induces effector recruitment\\
\hline
\hline
\multirow{2}{*}
{IL-2}          &  $alpha$, $p3$ and $g4$       &  Is produced   &\\
                \cline{2-4}
                &  $mu2$                        &  Is lost       &\\
\hline
\multirow{3}{*}
{TGF-$\beta$}  &  $p4$ and $tetha$             & Is produced &\\
               \cline{2-4}
               &  $mu3$                        & Is lost & \\
               \cline{2-4}
               &  $p2$ and $g3$                &         & Stimulates tumour growth\\

\hline
\end{tabular}
\label{Tab:ABSModelFourEquation}
\end{table}

\begin{table}[!htpb]
\centering
\caption{{\bf Transition rates calculations from the mathematical equations for case 3}}
\begin{tabular}{|l|l|l|l|}
\hline
Agent                &  Transition           & Mathematical equation                                         & Transition rate       \\
\hline \hline
\multirow{6}{*}
{Effector Cell}      & \multirow{3}{*}{Reproduce} & $\frac{p_1IE}{g1+I}\times$    & $\frac{p_1\times TotalIL\_2}{g1+TotalIL\_2}\times$\\

                     &                          &$\left ( p1 - \frac{q1S}{q2 + S} \right )$ & $\left ( p1 - \frac{q1 \times TotalTGFBeta}{q2 + TotalTGFBeta} \right )$\\

                     \cline{2-4}

                     & Die                       & $\mu_1E$                      & $mu1$\\

                     \cline{2-4}

                     & ProduceIL2                & $\frac{p3TE}{(g4 + T)(1+alphaS)}$         & $\frac{p3.TotalTumour}{(g4 + TotalTumour)(1+alpha.TotalTGF)}$\\

                     \cline{2-4}

                     & KillTumour                & $\frac{a_aTE}{g2+T}$  & $\frac{aa \times TotalTumour \times TotalEffector}{g2+TotalTumour}$\\

\hline
\hline
\multirow{5}{*}
{Tumour Cell}        &  Reproduce            & $\left ( aT \left (1- \frac{T}{1000000000}\right ) \right )$  & $\left ( TotalTumour.a \left( 1- \frac{TotalTumour}{1000000000}\right ) \right )$\\

                     \cline{2-4}

                     &  Die                  & $\left ( aT \left (1- \frac{T}{1000000000}\right ) \right )$  & $\left ( TotalTumour.a \left( 1- \frac{TotalTumour}{1000000000}\right ) \right )$\\

                     \cline{2-4}

                     &  DieKilledByEffector  & $\frac{a_a.TE}{g2+T}$              & message from effector\\

\cline{2-4}

                     & ProduceTGF & $\frac{p4T^2}{teta^2 + T^2}$                  & $\frac{p4.TumourCells}{teta^2 + TumourCells^2}$\\

                     \cline{2-4}

                     & EffectorRecruitment & $\frac{cT}{1 + \gamma S}$             & $\frac{c}{1 + gamma.TotaltGF}$\\

\hline
\hline

IL-2                & Loss                   & $\mu_2I$                             & $mu2$\\

\hline
\hline

\multirow{3}{*}
{TGF-$\beta$}         & Loss                   & $\mu_3S$                           & $mu3$\\

                    \cline{2-4}
                     & stimulates &&\\
                     &TumourGrowth             & $\frac{p2T}{g3+S}$                  & $\frac{p2.TotalTGF}{g3 + TotalTGF}$\\

\hline
\end{tabular}
\label{Tab:TumourEffector4TransitionRatesCalculations}
\end{table}

\begin{table}[!ht]
\centering
\caption{{\bf Parameter values for case 3}}
\begin{tabular}{|l|l|}
\hline
Parameter & Value\\
\hline
\hline
a         & 0.18\\
\hline
aa & 1\\
\hline
alpha     & 0.001\\
\hline
c & 0.035\\
\hline
g1 & 20000000\\
\hline
g2 & 100000\\
\hline
g3 & 20000000\\
\hline
g4 & 1000\\
\hline
gamma & 10\\
\hline
mu1 & 0.03 \\
\hline
mu2 & 10\\
\hline
mu3 & 10\\
\hline
p1 & 0.1245\\
\hline
p2 & 0.27\\
\hline
p3 & 5\\
\hline
p4 & 2.84\\
\hline
q1 & 10\\
\hline
q2 & 0.1121\\
\hline
theta & 1000000\\
\hline
k & 10000000000\\
\hline
\end{tabular}
 \label{Tab:CaseStudy2:ParametersModelCase3}
\end{table}

\clearpage

\begin{table}[!ht]
\centering
\caption{{\bf The results of the mixed model for the sequence of local maxima in case study 3}}

\begin{tabular}{|l|l|l|l|l|}
\hline
Technique & Parameter & Value & Std Error & p-value \\
\hline
\hline
ABMS & a & 0.1354 & 0.001403 & 0 \\
\hline
Gillespie & a &  0.2244 & 0.002242 & 0 \\
\hline
ABMS & b & 747.8501 & 2.876438 & 0 \\
\hline
Gillespie & b &  595.3515 & 3.338526 & 0 \\
\hline
\end{tabular}
\label{cs_3a}
\end{table}

\begin{table}[!hb]
\centering
\caption{{\bf The results of the mixed model for the sequence of local minima in case study 3}}
\begin{tabular}{|l|l|l|l|l|}
\hline
Technique & Parameter & Value & Std Error & p-value \\
\hline
\hline
ABMS & a & 0.01325 & 0.0002059 & 0 \\
\hline
Gillespie & a &  0.01823 & 0.0002514 & 0 \\
\hline
ABS & b & 37.33220 & 2.1334837 & 0 \\
\hline
Gillespie & b &  5.93249& 2.4013484 & 0 \\
\hline
\end{tabular}
\label{cs_3b}
\end{table}

\end{document}